\documentclass[aps,english,prl,tightenlines,reprint,11pt]{revtex4-1}

\usepackage{amsmath}    
\usepackage{amssymb}    
\usepackage{graphicx}   
\usepackage{verbatim}   
\usepackage{color}      
\usepackage[colorlinks=true,urlcolor=blue,citecolor=blue,linkcolor=blue]{hyperref}
\usepackage[all]{hypcap}
\usepackage{subcaption}
\usepackage{float}
\usepackage{braket}
\usepackage[dvipsnames]{xcolor}
\usepackage{caption}
\begin{document}
\title{DMRG Approach to Optimizing Two-Dimensional Tensor Networks}
\author{Katharine Hyatt}
\affiliation{Center for Computational Quantum Physics, Flatiron Institute, New York, NY 10010 USA}
\author{E.\ Miles Stoudenmire}
\affiliation{Center for Computational Quantum Physics, Flatiron Institute, New York, NY 10010 USA}
\date{\today}

\begin{abstract}
Tensor network algorithms have been remarkably successful solving a variety of problems in quantum many-body physics.
However, algorithms to optimize two-dimensional tensor networks known as PEPS lack many of the aspects that make the seminal density matrix renormalization group (DMRG) algorithm so powerful for optimizing one-dimensional matrix product states (MPS) tensor networks.
We implement an algorithm for optimizing PEPS which includes all of steps that make DMRG so successful for MPS. We show results for 2D spin models and discuss future extensions and applications.
\end{abstract}

\maketitle

\emph{Introduction}---Since the development of the density matrix renormalization group
algorithm (DMRG)~\cite{white1992density,White:1993a,White:2005,Schollwoeck:2005,mcculloch2008infinite} 
and its description in terms of matrix product states 
(MPS)~\cite{ostlund1995thermodynamic,dukelsky1998equivalence,schollwock2011density}, 
tensor networks have become a central technique for numerically studying quantum many-body
systems \cite{orus2014practical,Stoudenmire:2012a}. DMRG calculations have produced  
highly accurate results for challenging correlated electron physics
such as lattice models of spin-liquids \cite{LeBlanc:2015,Szasz:2018,HanQing:2019}, 
or ab-initio quantum chemistry calculations \cite{Kurashige:2013}.
The development of time-dependent and finite-temperature methods for MPS 
\cite{daley2004time,white2004real,Verstraete:2004m,Zwolak:2004,feiguin2005finite,white2009minimally}, 
has broadened the applicability of tensor network methods well beyond equilibrium, low-energy states. 
But most extensions of DMRG have focused on one-dimensional MPS tensor 
networks, which inherently do not scale to large two-dimensional systems. 

A promising ways to break through this scaling issue is to use  
two-dimensional tensor networks known as projected entangled pair states (PEPS)
which are a natural generalization of MPS \cite{Verstraete:2004p}. Not only can PEPS in principle capture 
low-energy states of two-dimensional systems, but they can be successfully optimized 
for challenging Hamiltonians.
Efficient algorithms have been devised to optimize PEPS using imaginary time evolution \cite{Jordan:2008}, gradient or conjugate gradient descent \cite{vanderstraeten2016gradient},
and variational optimization with a generalized eigenvalue solver \cite{corboz2016variational}.

But each of these methods lacks one or more of the essential steps
that has made the DMRG algorithm so powerful for optimizing MPS. 
Successfully realizing the high performance, precision, and reliability offered by DMRG but for 
two dimensional systems would enable rapid and definitive progress on many
pressing open problems in quantum many-body physics.

The specific technical aspects which make the DMRG algorithm so powerful are (1)~representing 
the Hamiltonian within an orthonormal basis during the entire calculation by 
enforcing a canonical form of the MPS; (2)~optimizing local tensors through 
solving a regular (not generalized) eigenvalue problem using fast, iterative solvers such
as Lanczos or Davidson; and (3)~the freedom to make large updates to individual tensors while maintaining overall stability. These aspects, taken together, allow DMRG to fully exploit the rapid convergence of 
iterative eigensolvers and translate this efficiency into optimizing the entire tensor network. For example,
in the very best cases of the DMRG algorithm for MPS, namely one-dimensional systems with an energy gap, one usually observes exponentially fast convergence of the energy as a function of DMRG sweeps or passes over the system.

In what follows, we demonstrate an algorithm for optimizing PEPS ground states which 
includes all three of the crucial technical aspects of DMRG listed above. 
Key ingredients in our work are recent advances in obtaining 
canonical forms for PEPS tensor networks and a description of their
properties \cite{Zaletel:2019,Hagshenas:2019}.
However, we take a different approach 
from Refs.~\onlinecite{Zaletel:2019} and \onlinecite{Hagshenas:2019} for carrying out the canonization at each step, raising the interesting question of which approach is best and underscoring how the efficiency and accuracy of  PEPS-DMRG methods will certainly continue to improve with further research.

\hypertarget{sec:mps}{\emph{Review of DMRG Algorithm for MPS}}---To set
the stage for discussing DMRG algorithm for optimizing 2D PEPS tensor networks,
let us first briefly review the original version of the algorithm for optimizing 
an MPS ground-state wavefunction, emphasizing key technical steps.
We will describe the one-site DMRG algorithm. Although the more common two-site DMRG algorithm
has the advantage of allowing the MPS bond dimension to be adjusted
adaptively, one-site DMRG is simpler to generalize
to our two dimensional setting.

\begin{figure} 
\includegraphics[width=0.8\linewidth]{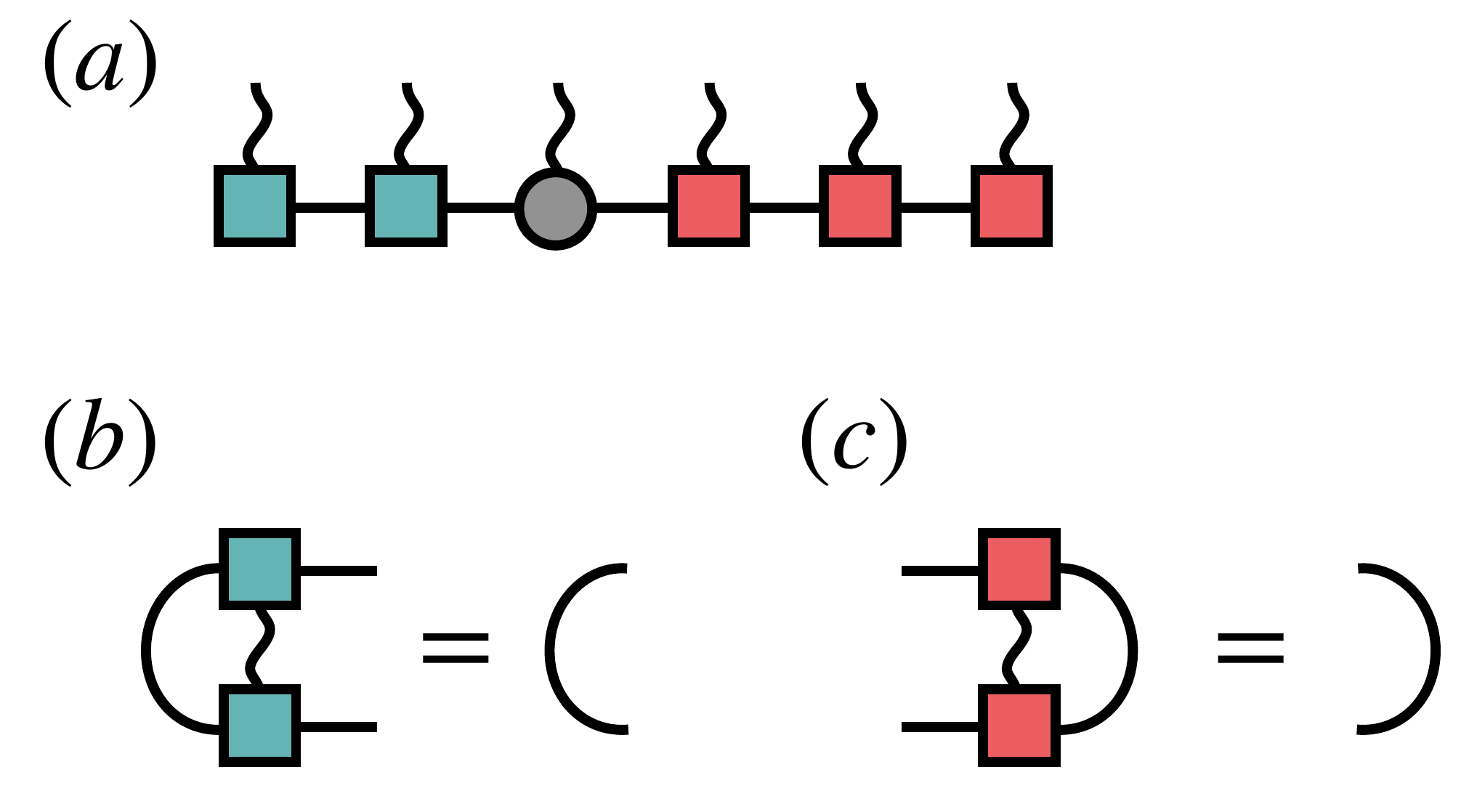}
\caption{Illustration of (a) a matrix product state (MPS) tensor network in canonical
form with respect to the third site, with tensors to the left obeying the (b) left-canonical
condition and tensors to the right obeying the (c) right-canonical condition. 
Curves without a shape attached to it represent identity operators in (b) and (c). 
Wavy lines represent physical or site indices of tensors.}
\label{fig:canonicalMPS}
\end{figure} 

The DMRG algorithm begins with an MPS in canonical form with respect to some site $j$
as illustrated in Fig.~\ref{fig:canonicalMPS}(a), with the figure showing the case \mbox{$j\!=\!3$}. 
By canonical form, one means all
MPS tensors to the left of site $j$ obey the left-orthogonality condition 
Fig.~\ref{fig:canonicalMPS}(b) and tensors right of $j$ obey the right-orthogonality
condition Fig.~\ref{fig:canonicalMPS}(c), where in each case squaring the tensor and contracting over two
of its indices results in an identity matrix. These orthogonality conditions turn out to
be crucial for the success of the DMRG algorithm, as each optimization step
can be written as a regular eigenvalue problem, making the algorithm numerically stable 
and efficient. In our PEPS algorithm, we will employ an analogous canonical form
for the same purposes.

To optimize the MPS within one step of the DMRG algorithm, only the center tensor at site $j$ 
is updated, holding the rest fixed. The terms of the Hamiltonian are projected into the orthonormal
basis defined by these fixed, canonical MPS tensors. This projection can be performed efficiently
by contracting the MPS tensors one at a time with each other and with the individual operators
making up the Hamiltonian.

After projection of the Hamiltonian terms into the MPS basis, one solves for the 
lowest energy eigenvector of the
projected Hamiltonian in order to obtain the updated tensor at site $j$. For efficiency, the eigenvalue
problem is solved using an iterative algorithm such as Lanczos. In practice, one does not
fully converge these solvers since the basis defined by the rest of the MPS is itself not fully optimized. The 
fact that DMRG optimizes each tensor this way is one of its key advantages: iterative eigensolver
algorithms converge very rapidly in practice, and because of the orthonormality of the MPS tensors,
local improvements translate directly into global improvements.

We will see that all of the steps above can be translated into the context of PEPS optimization. 
Some steps such as using an iterative eigensolver to optimize each tensor generalize
 straightforwardly from MPS to PEPS. Other steps such as canonization of tensors can 
 be done optimally for MPS, 
but are much more challenging for PEPS, and the question of which canonization algorithm 
is best remains an open question for future research.

For studying two-dimensional systems, MPS-DMRG can be applied to quasi-1D ladder-like 
($N_x \gg N_y$) lattices and scales linearly in $N_x$, but the bond dimension 
$D$ necessary to accurately represent a state must grow exponentially in 
$N_y$~\cite{yan2011spin,depenbrock2012nature}. Applying MPS-DMRG to 2D systems where 
$N_y$ is much greater than $N_y \approx 10$ while preserving a fixed accuracy is extremely expensive.
Thus a DMRG algorithm using PEPS as the underlying tensor network representation is highly desirable
as a way to make DMRG scale linearly in both $N_x$ and $N_y$.\\ 

\begin{figure}[H]
    \includegraphics[width=0.9\columnwidth]{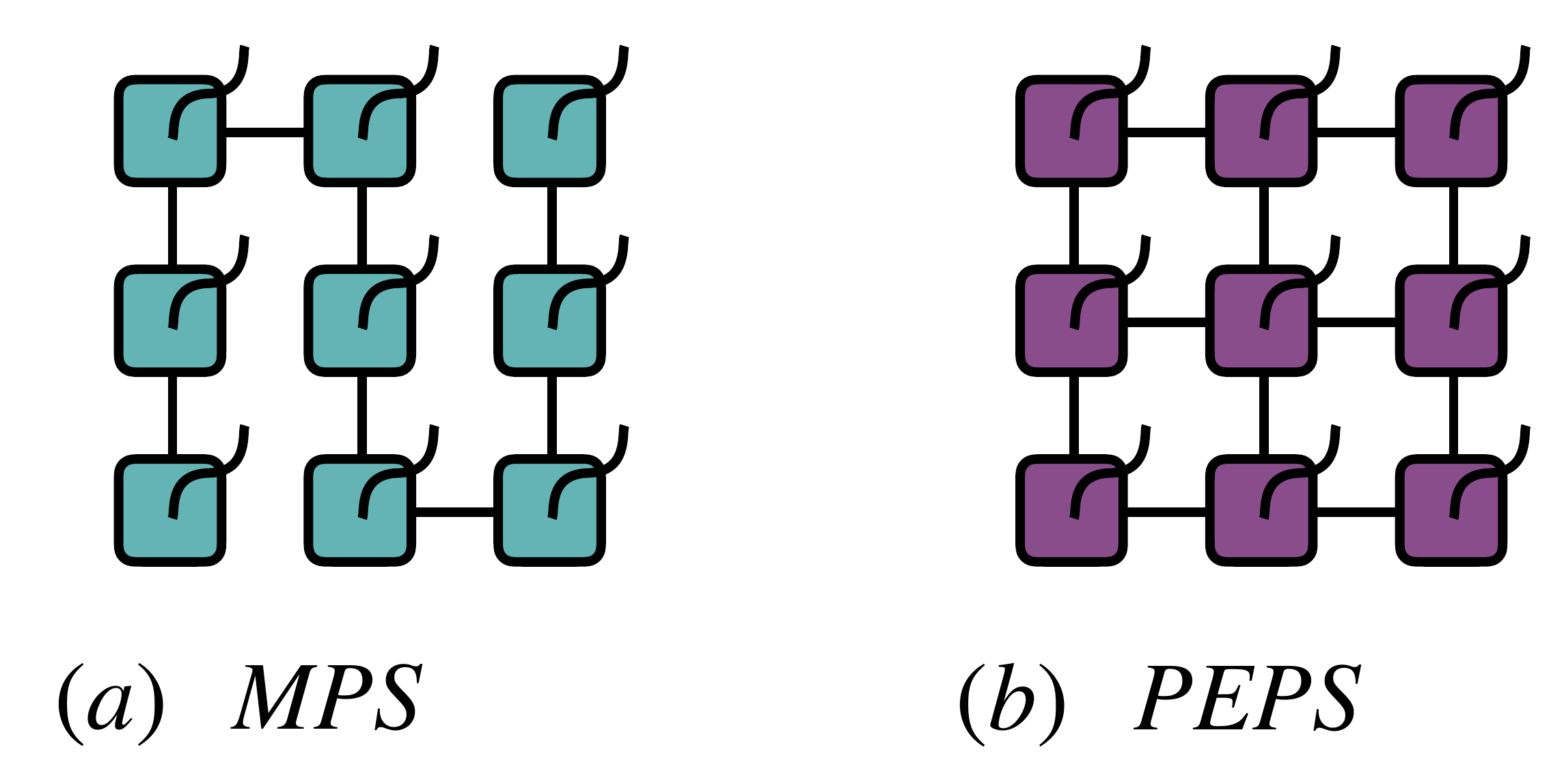}
    \caption{\small{\textbf{2D Tensor Networks:} (a) MPS with snaking pattern and 
    (b) PEPS, both shown for a $3 \times 3$ lattice with open boundaries.}}\label{fig:2DTNs}
\end{figure}

%
%
%
%
\hypertarget{sec:peps}{\emph{Projected Pair Entangled States}}---Projected pair entangled states (PEPS) 
\cite{Verstraete:2004p}
are a generalization of MPS tensor networks to two-dimensional systems.  
In an MPS, the virtual indices of each tensor connect to just two neighbors, resulting 
in a one-dimensional, chain-like network. 
PEPS tensors carry more virtual indices, typically chosen such that
the network of contracted virtual indices mimics the pattern of dominant interactions
in the Hamiltonian---see Fig.~\ref{fig:2DTNs}. 
Like MPS, each tensor in PEPS carries a \emph{physical} index for the physical degrees of freedom on the site. Unlike an MPS which requires an exponentially growing bond dimension
to capture ground states of two-dimensional systems of increasing size, 
PEPS are empirically known to capture ground states
using a modest bond dimension ($D \lesssim 20$) to a fixed accuracy 
independent of system size for a wide variety of Hamiltonians.

A key difference of PEPS from MPS is that computing observables of a PEPS, 
such as expected values of local operators, 
has an exponential cost if no approximations are used. This difficulty
can be traced back to the presence of loops in the contraction pattern of  
virtual PEPS tensor indices. Fortunately, observables can be computed accurately and reliably using various approximations.
In this work, observables are evaluated by a standard approach \cite{Jordan:2008} 
which treats the boundary columns
of the PEPS as MPS tensor networks, the interior columns as matrix product operator (MPO) networks, and 
uses MPO-MPS multiplication techniques to evaluate the contraction of the two, generating a new MPS which represents the contraction up to the first interior column.
This can be repeated to approximately contract the entire PEPS; see the Supplementary Material for more details.

A popular mode of using PEPS exploits translational invariance and
parameterizes the ground state of an infinite system by repeating a small unit cell
of PEPS tensors. The resulting infinite-PEPS (iPEPS) method \cite{Jordan:2008} then has the benefit 
of avoiding finite-size and boundary effects. However, there are important drawbacks too: 
iPEPS requires translation invariance and sophisticated algorithms must be developed to 
account for Hamiltonian terms acting outside of the particular tensor being optimized, 
while still ensuring translation invariance \cite{Jordan:2008,corboz2016variational,vanderstraeten2016gradient,jahromi2018infinite,Vanhecke:2019}.

Here we instead choose to work with finite PEPS with open boundary conditions.
Having an edge, and also not having to deal with translation-invariance constraints
provides a great technical simplification for the DMRG algorithm we will develop.

%
%
%
%
\hypertarget{sec:growing}{\emph{Canonization of PEPS}}---Unlike MPS with open 
boundary conditions, 
PEPS has internal closed loops which 
preclude an exact scheme in which the normalization matrix $\mathcal{N}$,
constructed by contracting all the tensors when computing the overlap $\langle \psi | \psi \rangle$ 
\emph{except} the tensors to be optimized, is equal to the identity.
However, DMRG-like schemes can be constructed which achieve $\mathcal{N} \simeq \mathbb{I}$ for a faithful representation of the PEPS with very small error~\cite{orus2014practical}.
One can also construct schemes in which $\mathcal{N} = \mathbb{I}$ exactly by incurring small
errors, or losses of fidelity of the PEPS during the canonization step. 
Here we develop the latter type of scheme, keeping in mind that it is not unique and others have been recently proposed \cite{Zaletel:2019,Hagshenas:2019}. 

\begin{figure}
    \hspace*{0.15in}
    \begin{subfigure}[c]{0.44\linewidth}
        \begin{centering}
        \includegraphics[width=0.9\linewidth]{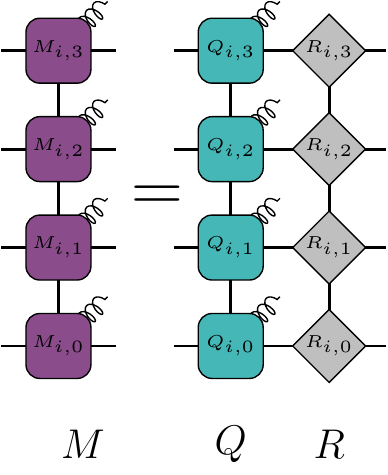}
        \end{centering} \caption{}\label{fig:gauge}
    \end{subfigure}
    \hspace*{0.2in}
    \begin{subfigure}[c]{0.34\linewidth}
        \begin{centering}
        \includegraphics[width=0.9\linewidth]{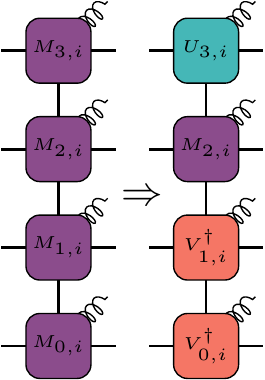}
        \vspace*{0.1in}
        \end{centering} \caption{}\label{fig:colgauge}
    \end{subfigure}

    \caption{\small{Canonization (a) of an entire PEPS column to shift to a different column  and (b) canonization within a specific PEPS column.}}\label{fig:canonization}
\end{figure}

As the simulation proceeds, the tensors in the column which was just optimized are 
``canonized'' before proceeding to optimize the next column. 
The canonization procedure takes as input a single column $M$ of the PEPS and outputs a new column $Q$ which is unitary and a non-unitary remainder $R$. The unitary piece $Q$ carries the physical indices and 
will replace the optimized column.
The non-unitary piece $R$ has the structure of an MPO and is multiplied into the next column 
to be optimized. Fig.~\ref{fig:canonization} illustrates the canonization procedure.
This procedure comes with a guarantee that $Q$ is always exactly canonical by construction, 
but the product of $Q$ times $R$ generally differs from  
the original, input column by a small amount. 
This discrepancy can be reduced mainly by increasing the number of optimization
passes over the $Q$ tensors and by enlarging the internal bond dimensions of $Q$ and $R$.

To briefly outline the canonization
procedure (see the Supplemental Material for more details):
first we make an initial guess for $Q$ by computing the polar decomposition
of each tensor of $M$ and using the unitary part of this decomposition to define each initial
tensor of $Q$.
Optimization of $Q$ then proceeds by computing $\text{Tr}[ M^\dag Q Q^\dag M]$, but leaving 
out one tensor of $Q$. The resulting tensor is called the environment tensor. 
One then computes a polar decomposition of this environment, whose unitary part gives an updated
$Q$ tensor which has maximal overlap with the environment \cite{Evenbly:2017,Vidal:2008,Schonemann:1966}. 
Having updated all $Q$ tensors this way, we compute $R=Q^\dag M$ using standard MPO multiplication
techniques and determine the inner product of $QR$ with $M$ to assess convergence.
Once the fidelity defined as $\text{Tr}[M^\dag Q R]/\text{Tr}[M^\dag M]$ is close enough to 1.0 
(typically $>0.99$ in our calculations) the 
canonization procedure is stopped and the algorithm proceeds to the next step.


Even after implementing the inter-column canonization above, finding the ground state 
would still require 
solving a generalized eigenvalue problem, $\hat{H}v - \lambda\hat{\mathcal{N}}v = 0$ 
($\hat{\mathcal{N}}$ is the normalization matrix).
To overcome this final obstacle we extend the scheme one step further.
One can also straightforwardly canonize \emph{within} a column, ensuring the 
correct normalization conditions above and below a specific tensor, given that the left and 
right environments are also properly canonized.
This procedure is depicted in Fig.~\ref{fig:colgauge}.
Just as in DMRG, this is done with an SVD or QR decomposition of the tensors above and below,
making this part of the canonization procedure optimal. 
This inter-column canonization ensures that the energy minimization can now be expressed as a \emph{regular} eigenvalue problem $\hat{H}v = \lambda v$, which is more efficient to solve and numerically stable. 
Just as in MPS-DMRG, we use an iterative eigensolver (specifically the Davidson algorithm) to optimize the local tensor,  rather than fully diagonalizing the Hamiltonian in the local basis. \\

%
%
%
%
\hypertarget{sec:heis}{\emph{Application to the Heisenberg Model}}---To test the reliability 
of the method, we compute the ground state of the spin 1/2 Heisenberg model on the square lattice, 
whose Hamiltonian is:
\begin{align} \hat{H} =& \sum_{\braket{i,j}} \hat{S}^x_i\hat{S}^x_j + \hat{S}^y_i\hat{S}_j^y + \hat{S}^z_i\hat{S}^z_j \label{eq:heis}\end{align}
where $\braket{i,j}$ means sites $i$ and $j$ are nearest-neighbors.
This model has also been extensively studied as a test-bed for PEPS algorithms~\cite{corboz2016variational,vanderstraeten2016gradient,bauer2009assessing,rader2018finite,lubasch2014algorithms,gu2008tensor}.

We study square lattice systems of size \mbox{$(L,L)$ with $L=4,6,8,10$} with open boundaries and apply
a staggered pinning magnetic field on the boundary spins.
The initial parameters in the PEPS are determined randomly, then the PEPS is optimized for 50 sweeps of DMRG, where a sweep means a pass over every PEPS tensor. 
We compute the energy per site and the nearest-neighbor spin-spin correlators throughout the simulation, comparing the energy with quantum Monte Carlo (QMC) and the correlators with MPS-DMRG. 
We allow the MPS-DMRG calculations to use a bond dimension up to 1000, since this value can get good accuracy for the Heisenberg model up through about $L \lesssim 12$. (Beyond a certain size $L$, MPS-DMRG with any fixed bond dimension starts to lose accuracy exponentially quickly.)

\begin{figure}
    \begin{centering}
    \includegraphics[width=1.0\columnwidth]{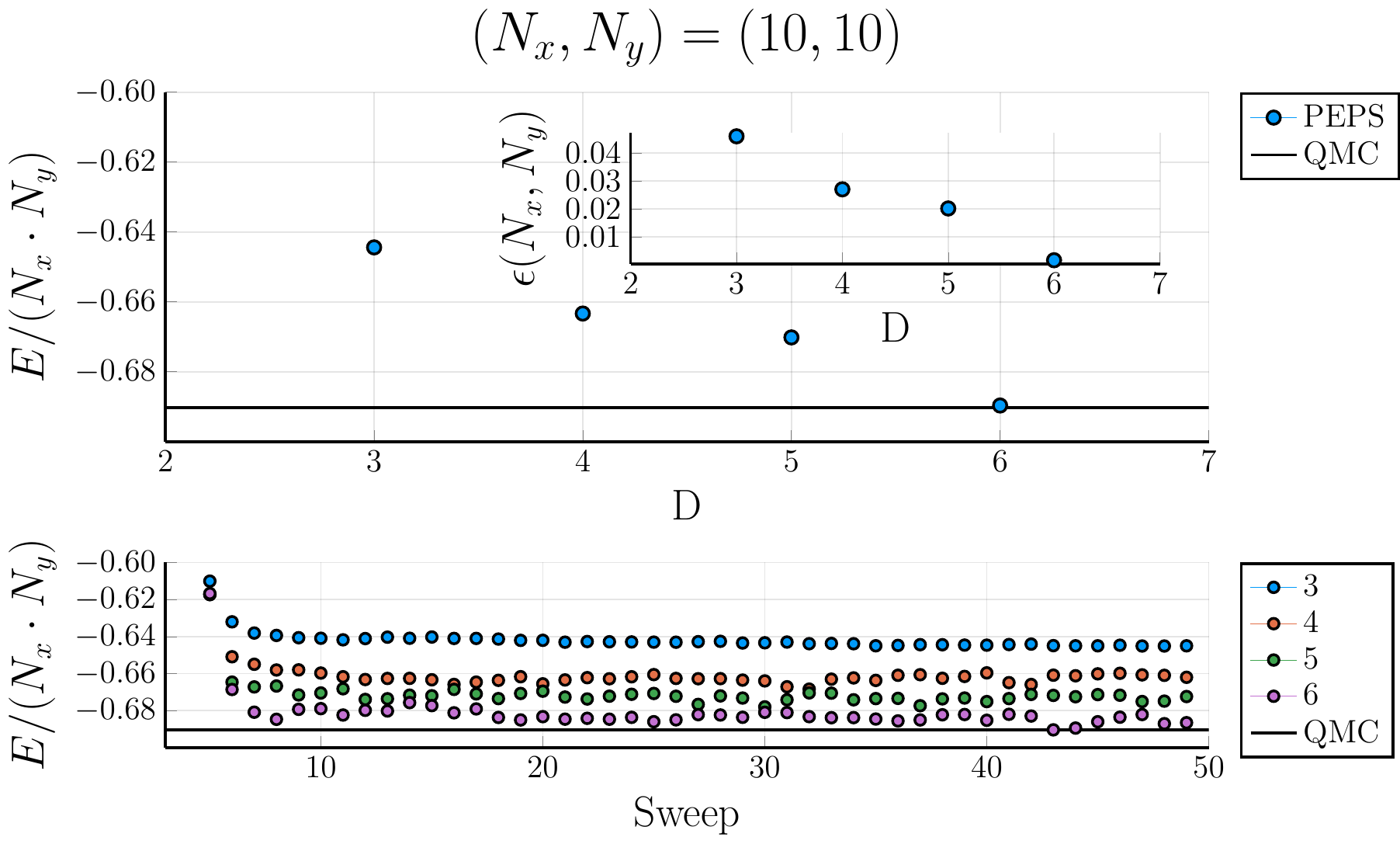}
    \end{centering}
    \caption{\small{Energy per site (top) $E/N$ for the Heisenberg model on a $10 \times 10$ finite square lattice with increasing PEPS bond dimension $D$. In addition to PEPS results, we show QMC results for the per-site energy. QMC results should be considered exact. The inset plot shows the difference between the PEPS-DMRG and QMC values. In the bottom subfigure we show $E/N$ as a function of sweeps over the system, compared with the exact QMC result.} \label{fig:heisenergy}}
\end{figure}

Progress towards the ground state is initially accelerated by performing simple update sweeps of the system for only the first several sweeps, before switching to eigensolver optimization
as in DMRG. Afterward, no more simple update steps are performed.

The energy results for PEPS-DMRG for systems of size $L=10$ are presented in the top plot of Fig.~\ref{fig:heisenergy},
with the inset showing the difference from the exact value.
The lower plot of Fig.~\ref{fig:heisenergy} shows the energy after each DMRG sweep over the system for 
the bond dimension $D=6$ PEPS. 
As expected, the convergence is quick in that only about ten sweeps are needed
to get within about $10^{-2}$ of the final energy.
One may ask why occasional increases in energy are seen: this occurs because the canonization process does not always converge to a sufficiently close approximation to the original non-canonical column, 
and in such cases replacing the column with its unitary approximation causes the energy to increase. 
This issue is not fundamental to the DMRG approach to optimizing PEPS,
and will improve with continuing progress on PEPS canonization algorithms.

\begin{figure}
    \includegraphics[width=0.9\columnwidth]{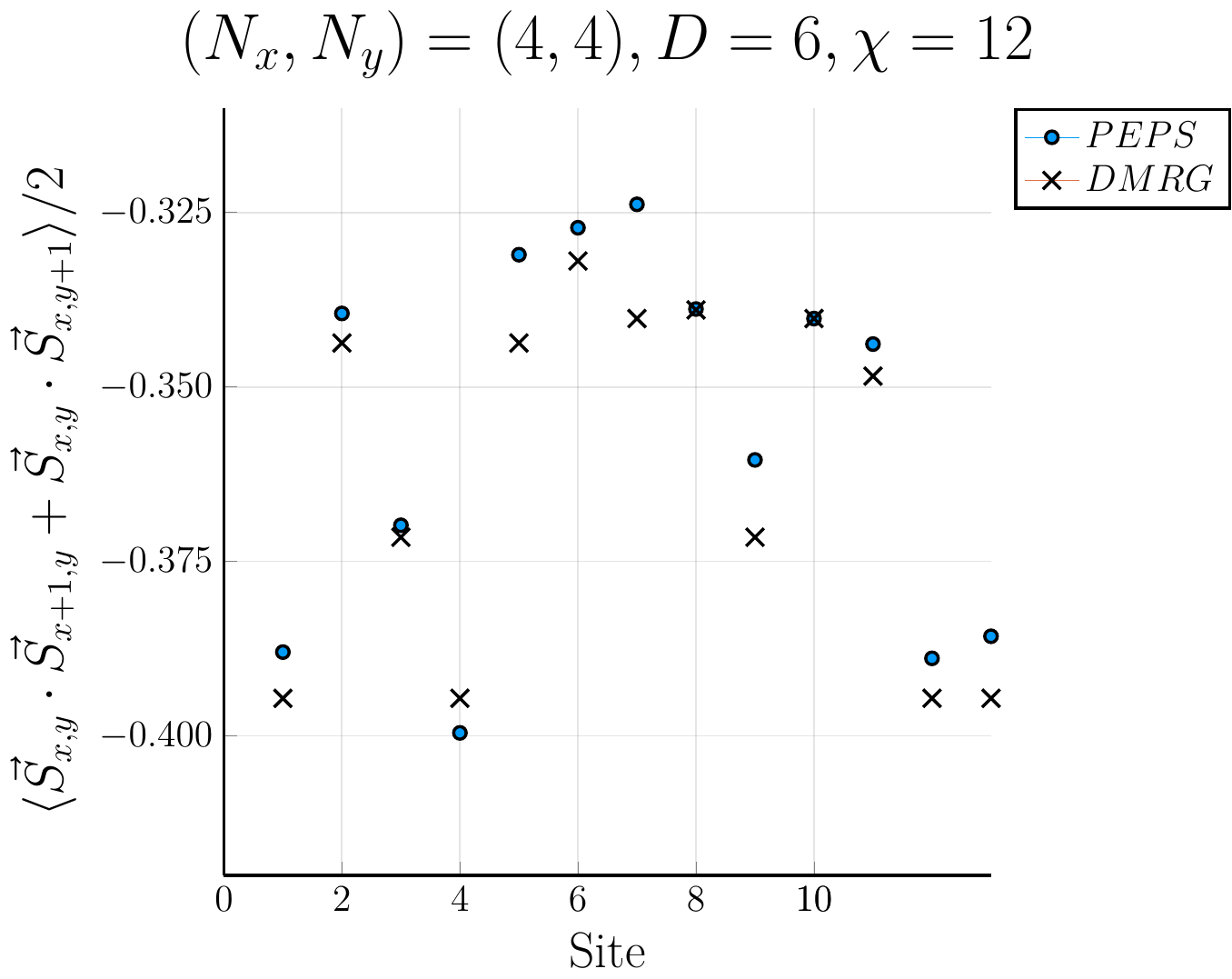}
    \caption{\small{Spin-spin correlators at each site of an $L=4$ Heisenberg model PEPS-DMRG calculation. The reference MPS-DMRG calculation ($\times$ symbols) is essentially exact for this system.}}\label{fig:heismag} 
\end{figure}

To further characterize the accuracy of our approach, we compute
 nearest-neighbor spin-spin correlators averaged between the horizontal and vertical directions, 
 that is 
 \mbox{$C_{x,y} = \langle \vec{S}_{x,y} \cdot \vec{S}_{x+1,y} + \vec{S}_{x,y} \cdot \vec{S}_{x,y+1} \rangle/2 $} and
compare with the exact results obtained using MPS-DMRG in Fig.~\ref{fig:heismag}.
The PEPS has bond dimension $D=6$ with an environment bond dimension of $\chi=12$.
Although there is a noticeable disparity between the PEPS results and the exact result, 
the difference is in large part a finite bond dimension effect and is comparable to that seen in state of the art iPEPS results~\cite{bauer2009assessing}.

%
%
%
%
\hypertarget{sec:conc}{\emph{Conclusions}}---We have demonstrated that the density matrix renormalization group (DMRG) algorithm, which has proven very powerful for variationally optimizing ground state wavefunctions in the MPS format, can be carried out successfully for PEPS
tensor networks, allowing one to scale DMRG to large two-dimensional systems. 
We have demonstrated that each essential step of DMRG can be performed with 
good results. 

Beyond applying the method to many open questions in 2D quantum systems,
there are also many readily attainable improvements to the algorithm. Of particular interest
would be a two-site version of the algorithm, giving the opportunity to grow
or shrink the PEPS bond dimensions adaptively. A two-site algorithm
should also perform much better than the one-site algorithm used here in avoiding
local minima. Scaling and performance optimizations are also
certainly within reach: we found the total run time is dominated by MPO-MPO multiplications (14\%) 
and by tensor contraction (46\%). Because the algorithm shown here is in large part built from
one-dimensional tensor network techniques, such as MPS-MPO products, 
improvements to these existing algorithms would be extremely helpful, 
and are an area where the community may be able to make significant progress.
In the course of this work we developed a GPU backend to the
ITensor library \cite{ITensor,ITensorsGPU} which significantly accelerated the numerical simulations, 
to be discussed further in a forthcoming work.
Last but not least, our approach would benefit greatly from a 
better understanding of methods to impose canonical or gauge conditions on PEPS tensors. 
Open challenges include determining the optimal bond dimensions after canonization 
and finding more efficient algorithms.

Looking ahead, it will be very interesting to compare the efficiency of optimizing 
finite PEPS with DMRG versus leading methods for optimizing infinite PEPS (iPEPS), 
in terms of estimating bulk properties. Note that the finite-size setting of the 
PEPS-DMRG algorithm we have implemented is not necessarily a drawback in comparison to iPEPS, 
as finite-size effects reveal useful physical information, and the finite setting allows
us to make very large updates to each tensor without encountering issues 
seen in iPEPS optimization relating to the non-linearity of optimization under 
translation-invariance constraints.

\begin{acknowledgments}
The authors thank Matthew Fishman, Michael Zaletel, Frank Pollman, Garnet Chan, Steve White, and 
Jing Chen for helpful discussions. All calculations are based on the ITensor Library~\cite{ITensor}. 
 The Flatiron Institute is a division of the Simons Foundation.
\end{acknowledgments}

\bibliography{peps_dmrg}

\end{document}